\documentclass[aps]{revtex4}
\usepackage{bm}
\usepackage{amsthm}
\usepackage{color}

\usepackage{graphicx,epsfig}
\usepackage[english]{babel}
\usepackage{amssymb,amsfonts,amsmath}
\usepackage{verbatim}

\def\bb#1{{\pmb{#1}}}

\def\+{\!+\!}
\def\-{\!-\!}
\def\={\!=\!}

\begin{document}

\title{Skyrmions and Hall viscosity}

\author{Bom Soo Kim}
\email{bkim2@loyola.edu}
\affiliation{Department of Physics, Loyola University Maryland \\
4501 N. Charles Street, Baltimore, MD 21210, USA }

\date{\today}

\begin{abstract}
We discuss the contribution of magnetic Skyrmions to the Hall viscosity and propose a simple way to identify it in experiments. The topological Skyrmion charge density has a distinct signature in the electric Hall conductivity that is identified in existing experimental data. In an electrically neutral system, the Skyrmion charge density is directly related to the thermal Hall conductivity. These results are direct consequences of the field theory Ward identities, which relate various physical quantities based on symmetries and have been previously applied to quantum Hall systems.  
\end{abstract}

\maketitle

\section{Introduction : Surprising relations between physical quantities} 

Recent progress in 2+1 dimensional hydrodynamics confirms that there exists a new hydrodynamic transport coefficient, Hall viscosity, in the absence of parity symmetry. This has provoked extensive theoretical investigations, yet experimental confirmation is still lacking. In particular, the Hall viscosity has been proposed to account for half of the angular momentum in systems with a mass gap, such as the integer and fractional quantum Hall systems and many more. Once one includes a background magnetic field, there is a similar relation between the Hall viscosity and Hall conductivity.  

We generalize previous discussions of Hall viscosity in the context of (non-)relativistic systems with(out) a mass gap. We then study the contribution to the Hall viscosity of magnetic Skyrmions. We derive various relations among transport coefficients and thermodynamic quantities in Skyrmion systems using Ward identities, parallel to the quantum Hall systems. One of these relations can explain experimental Hall conductivity data in the presence of Skyrmions. We propose a clear and simple way to measure the Hall viscosity in magnetic Skyrmion systems. We stress that ongoing Skyrmion experiments could be used to measure the Hall viscosity for the first time.   

To make this manuscript accessible to broader audience, we collect in the next section some basic facts about Skyrmions, review some recent developments, such as the appearance of the Skyrmion charge in the commutation relations of the stress energy tensor, and recall the definitions of the relevant transport coefficients and thermodynamic quantities. We then develop our Ward identities and apply them to Skyrmion physics. This review is based on results developed in \cite{Kim:2015qsa}.

\section{Background Materials}

We would like to introduce several different subjects whose underlying theme is parity breaking in 2+1 dimensions. These subjects have witnessed exciting developments in recent years: the theoretical understanding of topological Skyrmion charge as a central extension of commutation relations among energy momentum operators, experimental realizations of magnetic Skyrmions, and the role of Hall viscosity and angular momentum in hydrodynamics with broken parity. We will introduce these various concepts in a simplified context to gain familiarity and understanding, and consider more realistic scenarios with broken parity and boost symmetries later. In general, the larger the number of broken symmetries, the larger the number of transport coefficients that are expected to be present, and thus all the transport coefficients mentioned here will be present universally. Furthermore, when a symmetry is softly broken, the effects of the corresponding transport coefficients will be much smaller than the others.   

Finally, we review the Kubo formula connecting transport coefficients and retarded Green's functions. The Ward identities reveal their full power when the retarded Green's functions are rewritten in terms of transport coefficients. Along the way, we will find a natural way to incorporate thermodynamic quantities through so-called contact terms. In this way, we will be able to collect all relevant physical quantities in a systematic fashion.

\subsection{Skyrmion \& Central Extension}

One of the most striking features of quantum mechanics is the limitation of precise, simultaneous measurements of position and momentum. The uncertainty relation $ \Delta x^i \Delta p^j \geq \frac{\hbar}{2}\delta^{ij} $ originates from the commutation relations 
$[x^i, p^j] = i \hbar \delta^{ij} $, $[x^i,x^j]=0$, and $[p^i,p^j]=0$ where $i,j=1,\ldots,d$. We set $\hbar =1$. 

In some situations, the momentum commutation relation can be modified by a central extension. In particular, in some quantum field theories it is possible to have $ [P^i , P^j] = i C^{ij} $, where $ P^i = \int d^2 x ~T^{0i} (\vec x) $ is the momentum operator formed from the local energy momentum tensor $ T^{0i} (\vec x)$. There are several obstructions to having non-zero $C^{ij}$ due to Jacobi identities, which are necessarily satisfied. Let us start with the definition of the momentum operator that generates translations 
\begin{align}
[P^i,T^{0j}(t, \vec x)] = -i \partial_i T^{0j} (t, \vec x) \;.
\end{align}
Upon integrating this expression on both sides, we find that the right-hand side is a total derivative. To circumvent this, we need either a finite boundary contribution or a singularity in $T^{0j}$, which is realized in the later part of this section. The Jacobi identity involving a boost operator $M^{0i} $ and two momentum operators is violated by $ C^{ij} $:
$J(M^{0i}, P^0, P^j) = i C^{ij} $. In the presence of a boost symmetry, either Lorentz or Galilean, the left-hand side vanishes and thus $C^{ij}=0$. We may also consider the Jacobi identity involving a rotation operator $M^{ij} $ and two momentum operators. The contracted identity has the form $ \delta_{lj} J(M^{ij}, P^k, P^l) = i(d-2) C^{ki} $, where $d$ is the number of spatial dimensions. The left-hand side vanishes in the presence of rotation symmetry. Thus we see that, in the presence of rotational and translational symmetry, the Jacobi identity forbids a central extension except in two dimensions.  We will set $d=2$ henceforth.

Of course, there is a well known example for this type of construction: a constant background magnetic field with a modified energy momentum tensor $T^{0j}_B = T^{0j} - (B/2) \epsilon^j_n x^n J^0 $, where $J^0$ is a charge density operator. This modification corresponds to a minimally coupled momentum operator. Then 
\begin{align}
[P^i_B,T^{0j}_B(t, \vec x)] = -i \partial_i T^{0j}_B (t, \vec x) - i q \epsilon^{ijk} B^k J^0 (t, \vec x) \;.
\end{align}
There is a central extension due to the last term, for $J^0 \neq 0$. Boost symmetry is broken by the background magnetic field. 

A more interesting example with a central extension is a spin system with magnetic Skyrmions described by a continuous spin configuration $\vec n(t, \vec x)$, with $\vec n = (\sin \Theta (\rho) \cos \Phi (\rho), \sin \Theta (\rho)\sin \Phi (\rho), \cos \Theta (\rho) )$ in the coordinate $\vec x = (\rho \cos \phi, \rho \sin \phi, z)$. The Lagrangian density is given by
\begin{align}
\mathcal L = \dot \Phi (\cos \Theta - 1) - (J/2) \partial_i \vec n \cdot \partial_i \vec n \;,  \qquad [\Phi (t, \vec x), p_{\Phi} (t, \vec x')] = i \delta^2 (\vec x - \vec x') \;,
\end{align}
where $p_{\Phi} (t, \vec x) = \cos \Theta (t, \vec x) -1 $. Then \cite{Watanabe:2014pea}\cite{Toma1991}
\begin{align}\label{CommutationRelationC}
[P^i,T^{0j}(t, \vec x)] = -i \partial_i T^{0j} (t, \vec x) + i \epsilon^{ij} (\epsilon^{kl} p_\Phi (t, \vec x) \partial_k \partial_l \Phi (t, \vec x) )\;.
\end{align}
The energy momentum tensor is given by $ T^{0i} (t, \vec x) = p_\Phi (t, \vec x) \partial_i \Phi (t, \vec x) $. 
This system has a central extension in the commutation relations involved in the energy momentum operators because the field $\Phi$ has a vortex singularity at $\Theta = \pi$. The singularity also breaks boost symmetry. The conservation equation $ \partial_\mu T^{\mu\nu}=0$, with $\mu, \nu = 0,1,2$, can be shown to be satisfied explicitly. 

It will be convenient to use a local version of \eqref{CommutationRelationC}
\begin{align}\label{CommutationRelationLocal}
[T^{0i}(t, \vec x),T^{0j}(t, \vec x')] = i \left(- \partial_i T^{0j} + \partial_j T^{0i} + i \epsilon^{ij} c \right) \delta^2 (\vec x - \vec x') \;.
\end{align}
While we have presented this relation in the context of a particular spin model, it holds for Skyrmion systems in general, independent of the details such as the form of Lagrangian given above. 

The total Skyrmion charge is the integral of the charge density $c$. Let us consider a spin configuration that is composed of up-spins $\uparrow$ at infinity and down-spins $\downarrow$ at the center. Then 
\begin{align}
C &=\! \int d^2 x ~\vec n \cdot \left[\frac{\partial \vec n}{\partial x} \times \frac{\partial \vec n}{\partial y} \right] 
= \int_0^\infty \!\! d\rho \int_0^{2\pi} \!\! d\phi \frac{d \Theta}{d\rho} \frac{d \Phi}{d\phi}	\sin \Theta
= \cos \Theta (\rho) \Big|_{0}^{\infty} \Phi(\phi) \Big|_{0}^{2\pi} 
= 2 \cdot 2\pi m.
\end{align} 
Here $m$ is the number of $2\pi$ rotations of the spin between the infinity and the center. Only $m=\pm 1$ are stable.  
Magnetic Skyrmions of this sort have been first observed in MnSi using neutron scattering \cite{SkyrmionExp1}. Skyrmion spin structures have also been observed with a real space Lorentz transmission electron microscope in FeCoSi thin films \cite{SkyrmionExp2}. Since then Skyrmions have been observed in many different materials \cite{Review}.

\subsection{Parity breaking hydrodynamics in 2+1 dimensions}

Hydrodynamics is an effective theory, describing dynamics at large distance and time scales. It incorporates dissipative effects. Thus it can be described at best by conservation equations. The central object is the energy momentum tensor $T^{\mu\nu}$. In relativistic hydrodynamics, the hydrodynamic equations are $\partial_\mu T^{\mu\nu} = 0$. The variables are temperature $T$ and velocity $u_\mu$ with $u^\mu = (u^0, u^i)$. If we normalize the velocity as $ \eta_{\mu\nu} u^\mu u^\nu = -1$, there are equal number of equations and variables. Thus one can solve the problem. 

We assume local thermal equilibrium and solve order by order in the derivative expansion. In the derivative expansion, there are redundancies, and we fix them with the Landau frame condition, $T^{\mu\nu} u_\nu = - \varepsilon u^\mu$, which says that the flow of the fluid is equal to the energy flow. We further impose the local second law of thermodynamics, which can be described by the equation $\partial_\mu T^{\mu\nu}u_\nu = 0$. This leads to the so-called entropy current, which is required to be positive definite, which in turn will put constraints on various transport coefficients. 

If we impose Lorentz invariance, there are only two quantities allowed in the energy momentum tensor, energy and pressure, at leading (ideal) order. At the first derivative order, we have another two, the shear and bulk viscosity, multiplying the shear tensor $\sigma^{\mu\nu}$ and the divergence of the velocity, respectively. Combining them together, 
\begin{align}
T^{\mu\nu}(T,u)  & = \varepsilon  u^\mu u^\nu + p P^{\mu\nu}\-\eta \sigma^{\mu\nu} 
- \zeta P^{\mu\nu}  (\partial_\alpha u^\alpha) \;, 
\end{align}
where the shear tensor is $\sigma^{\mu\nu} =P^{\mu\alpha} P^{\nu\beta} \big[\partial_\alpha u_\beta + \partial_\beta u_\alpha - P_{\alpha\beta} (\partial_\delta u^\delta) \big]$ and the projection operator is $ P^{\mu\nu} = u^\mu u^\nu + \eta^{\mu\nu}$. This is the relativistic hydrodynamics of Landau and Lifshitz. 

It turns out that we can create two new tensors using the epsilon tensor if we allow parity breaking effects in 2+1 dimensions. Starting from the shear tensor we can create $\tilde{\sigma}^{\mu\nu}  = \epsilon^{\alpha\beta(\mu}u_{\alpha} \sigma_{\beta}^{\ \ \nu)}$; the corresponding transport coefficient is called Hall viscosity $\eta_H$, which we will review below. By differentiating the velocity, we can generate another term called the vorticity $\tilde \Omega = - \epsilon^{\beta\gamma\alpha} u_{\beta}\nabla_{\gamma}u_{\alpha}$. If one subtracts the vorticity from the energy momentum tensor, in the so-called vortical-frame, then the coefficient is nothing but the Hall analogue of the bulk viscosity. Thus we can add two more terms if the parity is broken! This has been developed recently \cite{Jensen:2011xb}\cite{Bhattacharya:2011tra}. Combining all the terms, we have 
\begin{align}
T^{\mu\nu}(T,u)  &= \varepsilon  u^\mu u^\nu + p P^{\mu\nu}-\eta \sigma^{\mu\nu} 
-\zeta P^{\mu\nu}  (\partial_\alpha u^\alpha) 
-\eta_{H} \tilde{\sigma}^{\mu\nu} -\chi_{H} P^{\mu\nu} \tilde \Omega  \;.
\end{align}
This can be extended for the charged case, which will be slightly more complicated. If boost symmetry is broken, there can be more transport coefficients on top of the transport coefficients that have been mentioned above \cite{Hoyos:2013eza}. We would like to point out that these transport coefficients can be directly related to each other and/or to thermodynamic quantities. 

The purpose of this exercise is to make sure that we do not miss universally present hydrodynamic quantities for physically interesting systems such as Skyrmions.

\subsection{Hall viscosity} 

We study Hall viscosity in a slightly different setting to explain its physical meaning more clearly. Let us focus on our discussion for fluids.  A small deformation described by $\xi_i$ produces a stress $T_{ij}$ depending on strain that is described by $\xi_{ij} = \partial_i \xi_j + \partial_j \xi_i$ and the strain rate $\dot \xi_{ij} = \partial_t \xi_{ij}$. Then the stress-energy-momentum tensor has the form   
\begin{align}
T_{ij} & = p \delta_{ij} - \lambda_{ijkl}\xi_{kl} 
- \eta_{ijkl} \dot \xi_{kl}  \;. 
\end{align} 
For rotationally invariant systems, we can constrain the transport coefficients in terms of the elastic modulus $\lambda$ and viscosity tensors $\eta, \zeta$  as  
\begin{align}
\lambda_{ijkl} = \lambda \delta_{ij}\delta_{kl}= -V \frac{\partial p}{\partial V} \delta_{ij}\delta_{kl} \;, \quad
\eta_{ijkl} = \eta (\delta_{ik}\delta_{jl}+\delta_{il}\delta_{jk}) 
+(\zeta-\eta)\delta_{ij}\delta_{kl} \;.
\end{align} 

For systems with broken parity, for example in the presence of a background magnetic field, the odd part of the shear tensor is allowed \cite{Avron:1995}. Explicitly, 
\begin{align}
\eta_{ijkl}^A = -\eta_{klij}^A =-\frac{\eta_H}{2} 
(\epsilon_{ik}\delta_{jl}+\epsilon_{jl}\delta_{ik} +\epsilon_{il}\delta_{jk} + 
\epsilon_{jk}\delta_{il}) \;.  
\end{align} 
For definiteness, we consider two spatial dimensions $d=2$ with indices $i,j=1,2$. One can explicitly check that this quantity is symmetric under the exchange of the indices $i$ and $j$ as well as $k$ and $l$. On the other hand, it is anti-symmetric under the exchange of $ij$ and $kl$. 

To reveal the physical meaning of the Hall viscosity, let us imagine a fluid containing a finite size cylinder in the middle rotating counterclockwise with constant frequency. The well known shear viscosity acts as a clockwise force along the surface of the cylinder trying to slow down the cylinder. On the other hand, Hall viscosity acts perpendicular to the surface of the cylinder, outward or inward depending on the situation. Thus Hall viscosity has nothing to do with dissipation of the motion of the cylinder.   

Let us consider the energy change under a small deformation of the strain. This is described by $ \delta \varepsilon = - T_{ij} \delta \xi_{ij}$. Then $T\dot s =  \eta_{ijkl} \dot \xi_{ij} \dot \xi_{kl}  + \eta^A_{ijkl} \dot \xi_{ij} \dot \xi_{kl}$, where $s$ is the entropy density. It turns out to be positive semi-definite, and we can use this to put some bounds on the shear and bulk viscosities. Now, the second term $\eta^A_{ijkl} \dot \xi_{ij} \dot \xi_{kl}$ actually vanishes because the strain rates are symmetric under the exchange of the indices $ij$ and $kl$, while $\eta^A_{ijkl}$ is anti-symmetric. Thus $\eta_H $ is dissipationless and can exist even at zero temperature. The Hall viscosity has a better chance to be observed in low temperature experiments, where dissipative effects become small.

\subsection{Angular momentum} 

Let us now review angular momentum, which also plays role in the Ward identities. We consider, for simplicity, a static case described by the conservation equation $ \partial_i T^{0i}(\vec x) = 0$. This equation has an obvious solution 
\begin{align}
\langle T^{0i}(\vec x) \rangle = \epsilon^{ij} \partial_j \ell (\vec x) \;, \qquad \qquad
\ell (\vec x) =\begin{cases} 
\frac{1}{2} \ell & \quad(|x|, |y| \leqslant b)\\
0 & \quad(\text{otherwise}) \end{cases}\;,
\end{align}
where $\ell$ is a constant. So $T^{0i} (\vec{x})$ is vanishing both inside and outside the confined region (a square with side $b$), and is only non-vanishing at the boundary. 
\begin{align}
T^{0i}(\vec{x}) =\frac{1}{2} \ell \epsilon^{ij} \left[ -\delta(x^{j} - \frac{1}{2}b) + \delta(x^{i} + \frac{1}{2}b)\right] \theta(\frac{1}{2}b-|x|) \theta(\frac{1}{2}b-|y|) \;.
\end{align}
This corresponds to a momentum flow around the edge of the square box, an edge current. $\ell$ characterizes its strength. The direction of the edge current is always along the edge, either clockwise or counter-clockwise, depending on the sign of $\ell$. This boundary nature is related to the topological nature of the underlying field theory.

Now we consider the infinite volume limit. Normally we discard the effects of $ T^{0i}$ because it is a boundary contribution. However, a careful treatment show that it indeed makes a contribution to the total angular momentum \cite{Liu:2012zm}:
\begin{align}
L = \int d^2 \vec x \epsilon_{ij} x^i \langle T^{0j} \rangle 
= \int d^2\vec x\ell(\vec x)\partial_i x^i=\ell\int d^2\vec x=\ell V_2 \;.
\end{align} 
The total angular momentum is independent of the shape of the boundary. This tells us that identifying the angular momentum is subtle. The above example illustrates that spontaneously generated angular momentum even in a system without boundary can play important role and needs to be considered carefully.  

We would like to point out some even more subtle points related to symmetries. It turns out that we cannot keep both translation and rotation invariance in the presence of the expectation value of the energy momentum tensor $\langle T^{0i}(\vec x) \rangle \neq 0$. (This is also true in the presence of boundary. We do not consider the boundary effects in the rest of this paper.) Thus we have two choices: we can either keep both translation and rotation invariance without angular momentum or keep rotation invariance and angular momentum without translation invariance. These two different choices lead to two highly non-trivial and mutually exclusive sets of Ward identities. 

\subsection{Kubo formula} 

To capture all the known transport coefficients and thermodynamic quantities that describe a certain system, we couple the system to external sources, such as background gauge fields $A_i$ and background metrics $g_{ij} $, compute the variation, and eventually set the sources to zero. The corresponding currents and energy momentum tensors can be obtained from the variation of partition function ${\mathcal Z}[A_i , g_{ij} ]$ by employing linear response theory and expanding around $ A_i=0, g_{ij}=0$.  
\begin{align}
\begin{split}
J_i &= \sigma_{ij} E_j = \frac{\delta \log \mathcal Z}{\delta A^i} 
= - \bar n A_i + \sigma_{ij} E_j + \cdots \\
T_{ij} & = \frac{2}{\sqrt{-g}} \frac{\delta \log \mathcal Z}{\delta g^{ij}}  
= p \delta_{ij} -\frac{\kappa^{-1}}{2} \delta_{ij}\delta_{kl} h_{kl} 
-\frac{1}{2} \eta_{ijkl} \dot h_{kl}  + \cdots \;,
\end{split}
\end{align} 
where $\bar n$ is a charge density, $E_i = \partial_t A_i - \partial_i A_t$ is the electric field, $\sigma$ is the conductivity, $p$ is the pressure, $\kappa$ is the inverse compressibility, and $\eta$ is the shear viscosity we defined previously. The first term on the right side of the first equation (and the first two in the second one) is a diamagnetic term. This is a contact term and can be obtained by including an $A^2$ term in the partition function. 

Now $\sigma_{ij}$, one of the transport coefficients, can be obtained by taking another variation of $J_i$ with respect to $A_j$, and then setting $A_i=0$.  Then in the momentum space, 
\begin{align}
\begin{split}
\sigma_{ij} (\omega, {\bf q}) &= -i \frac{\bar n}{\omega^+} \delta_{ij} 
+ \frac{i}{\omega^+} G^R_{ij} (\omega, {\bf q}) \;, \\
G^R_{ij} (\omega, {\bf q}) & = \int dt\int d^d x e^{i\omega^+ t - i {\bf q} \cdot {\bf x}}
~i \theta (t) ~ \langle [J_i (t, {\bf x}), J_j (0, {\bf 0})]\rangle \;. 
\end{split}
\end{align} 
The factors $1/\omega$ come from time derivative in $E_i$ and the Fourier transform to the momentum space. Thus Kubo formulae relate transport coefficients to retarded Green's functions. Note that differences between them arise from different available contact terms. There is a similar relation between $\eta_{ijkl}$ and $g_{ij}$.  In the previous literature, these contact terms had been missed.

It turns out that we do not need to go through this general exercise. Our Ward identity is powerful enough to capture all the possible terms including the contact terms. We double-check our Ward identity results with available results when possible.

\section{Ward identities \& Skyrmions}

In this main section, we make use of the Ward identities for Skyrmions to provide various relations among the transport and thermodynamic quantities, that have been reviewed in the previous section. First, we provide a simple example that illustrates the physical meaning of the Ward identity. By considering the translation and rotation invariant case, we provide physical meaning to the topological charge $c$ in the context of transport coefficients in \S \ref{sec:WIInsulating}, along with a simple and clear way to observe the Hall viscosity in insulating Skyrmion materials. We generalize our Ward identities with charge in the presence of a magnetic field in \S \ref{sec:WIChargeB}. In \S \ref{sec:HallConductivity}, we explain the implications of the Ward identities for the phenomenology of the Hall conductivity and its connection to experiments. We further provide a simple formula for the Hall viscosity in the charged case in terms of Hall conductivity measurements as a function of momentum in \S \ref{sec:HallViscosityConductor}. Finally, we briefly comment on the case  without translation invariance in \S \ref{WIAngularMomentum}.   

\subsection{Geometric understanding of Ward identities} 

Let us consider a simple and geometric picture that illustrates the physical meaning of the Ward identities. In 2 spatial dimensions, there are only two independent area preserving shear transformations. For a square, we can (a) elongate one side and squeeze the other or (b) stretch along the diagonal direction. In terms of matrices, we can represent these transformations  as   
\[ a = \left( \begin{array}{cc}
1 + \epsilon & 0  \\
0 & 1 - \epsilon \end{array} \right), \quad 
b = \left( \begin{array}{cc}
1 & \epsilon'  \\
\epsilon' & 1 \end{array} \right); \qquad\quad  
b^{-1} a^{-1} b a =  \left( \begin{array}{cc}
1 & -2\epsilon \epsilon'  \\
2\epsilon \epsilon' & 1 \end{array} \right) + 
\mathcal O(\epsilon, {\epsilon'})^3 \;.
\] 
It turns out that $a$ and $b$ do not commute each other, but the operation $b^{-1} a^{-1} b a$ produces a net rotation. Thus shear transformations can generate rotations! 

Let us promote this observation to the level of a quantum field theory Ward identity. What we will get is a rather non-trivial relation between the Hall viscosity and angular momentum:  $\eta_H = -\frac{1}{2} {\ell} $, meaning that the Hall viscosity is half of the angular momentum $\ell$ of the system \cite{Read:2008rn}\cite{Read:2010epa}. This has been theoretically demonstrated for various integer and fractional quantum Hall states.  
On the other hand, if one includes a background magnetic field ($B\neq 0$) with Galilean invariance, one can obtain another non-trivial relation stating that Hall viscosity $\eta_H$ is proportional to Hall conductivity $\sigma_H$ \cite{Hoyos:2011ez}. 

Now, how can we get two different results for the same quantity, the Hall viscosity? It turns out that they are two relations depending on the choices of the symmetries discussed in the angular momentum section: keeping rotation invariance and angular momentum without translation invariance or keeping both translation and rotation invariance without angular momentum. These two choices are incompatible as mentioned and lead to two independent results \cite{Bradlyn:2012ea}\cite{Hoyos:2014lla}\cite{Hoyos:2015yna}.

\subsection{Simple Ward identities}

Let us examine a simple Ward identity by taking two time derivatives of a retarded Green's function 
$G^{0j0l}(x^\mu ;x'^{\mu} )=  i\Theta(x^0-x'^{0}) \langle [ T^{0j}(x^\mu ), T^{0l}(x'^{\mu} )] \rangle $, where $ \Theta(x^0-x'^0)$ is a step function. There are 4 terms, with the two derivatives acting on the two $T$s, the step function, or both. They are organized as   
\begin{align}
{\partial_0' \partial_0 G^{0j0l}(x^\mu ;x'^{\mu}) }
& = \partial_n \partial_m' G^{njml}(x^\mu ;x'^{\mu})	
- \left[ \delta' (x^0-x'^0) + \delta (x^0-x'^0) (\partial_0 - \partial_{0}')\right] 
C^{0j0l}(x^\mu ;x'^{\mu}) \;. 
\end{align}  
The conservation equation $\partial_\mu T^{\mu j} = \partial_0 T^{0j} + \partial_m T^{m j}= 0$ is used to rewrite the first term. The other three terms are contact terms that carry the delta function, the derivative of delta function, and $C^{0j0l}(x^0,\vec x ,{\vec x}' )=  i\langle [ T^{0j}(x^0,\vec x ), T^{0l}(x^0,{\vec x}' )] \rangle$ defined at the same time due to the delta function. 

Now the key observation is the fact that the contact terms can be rewritten as one point functions using \eqref{CommutationRelationLocal}. 
Further assuming time translation symmetry and performing a Fourier transform in the time direction as $\int d(x^0\- x'^0)e^{i\omega (x^0 - x'^0)} \cdots $, we arrive at our general result in this simplest case.  
\begin{align}\label{GeneralWI}
\begin{split}
\omega^2 G^{0j0l} (\omega,\vec x ,{\vec x}' ) 
&= \partial_n \partial_m' G^{njml}(x^\mu ;x'^{\mu} ) \\
&+\frac{1}{2}\left[2i\omega \epsilon^{jl} \epsilon^n_{\ m} \partial_n \langle T^{0m}(\vec x ) \rangle 
- \partial_l \partial_n \langle T^{nj} (\vec x ) \rangle
-  \partial_j \partial_m \langle T^{ml} (\vec x ) \rangle  \right] \delta(\vec x -{\vec x}' )  \\
&~{- i \omega c^{jl} \delta(\vec x -{\vec x}' )} \;.
\end{split}
\end{align} 
Ward identities are actually consequences of the conservation equation $\partial_\mu T^{\mu j} = 0$. Here we further generalize to include the central extension that arises from the commutation relations among energy momentum operators! This is our new result \cite{Kim:2015qsa} based on \cite{Hoyos:2014lla}\cite{Hoyos:2015yna}. 

As advertised before, there are two independent and exclusive cases we can consider depending on symmetries. We consider one at a time. 

\subsection{With rotation and translation invariance} 

Here we impose rotation and translation symmetries and examine the consequences of the Ward identities. With these symmetries, all contact terms vanish because translation invariance does not allow any spatial dependence for the one point functions! Thus \eqref{GeneralWI} becomes 
\begin{align}\label{TranslationWI}
&\omega^2  G^{0j0l} (\omega,\vec x ,{\vec x}' ) 
= \partial_n \partial_m' G^{njml}(x^\mu ;x'^{\mu}) 
- i \omega c^{jl} \delta(\vec x -{\vec x}' ) \;.
\end{align}  
We use translation invariance to perform a Fourier transform $G^{\mu\nu\alpha\beta}(\omega,\vec x - {\vec x}')$ $= 1/(4\pi^2) \int d^2 \vec q$ $  e^{i\vec q\cdot (\vec x -{\vec x}' )} $  $\tilde G^{\mu\nu\alpha\beta}(\omega,q)$. We then use rotation invariance to write down the most general tensor structures for the retarded Green's functions:
\begin{align}\label{IndexStructures}
\begin{split}
&\tilde G^{0i0k} = -i\omega \big[\delta^{ik} \bm{\kappa}_\delta +\epsilon^{ik} \bm{\kappa}_{\epsilon}
+ q^i q^k\bm{\kappa}_{q} +(\epsilon^{in}q_n q^k+\epsilon^{kn}q_n q^i) \bm{\kappa}_{q\epsilon} \big]\;, \\
&\tilde G^{njml} 
= -i\omega \Big[\eta(\delta^{nm}\delta^{jl}+\delta^{nl}\delta^{mj} -\delta^{nj}\delta^{ml})
+\zeta\delta^{nj}\delta^{ml} 
+\frac{\eta_H}{2}(\epsilon^{nm}\delta^{jl}+\epsilon^{nl}\delta^{jm}+\epsilon^{jm}\delta^{nl}
+\epsilon^{jl}\delta^{nm})\Big].
\end{split}
\end{align}  
We already saw a similar tensor structure with rotation invariance. The parts with the epsilon tensors are due to the absence of parity invariance. We identify $\bm{\kappa}$ as the thermal conductivity. 
The proper name might be momentum conductivity which is related to the response of the momentum density and current, $T^{0i}$ and $T^{ij}$! 

After a little algebra, we get 
\begin{align}\label{MomWI2}
&\omega^2  \left[  \delta^{jl}\bm{\kappa}_\delta + \epsilon^{jl} \bm{\kappa}_{\epsilon}
+ q^j q^l \bm{\kappa}_{q}
+ (\epsilon^{jn}q_n q^l+\epsilon^{ln}q_n q^j) \bm{\kappa}_{q\epsilon} \right]
= \delta^{jl} q^2\eta  + q^j q^l \zeta + \epsilon^{jl} ({c}+ q^2 \eta_H ) \;. 
\end{align} 
This equation looks like it should give a single Ward identity. Yet there are 4 independent tensor structures, and thus there are 4 independent Ward identities depending on the tensor structures. Applying this to the parity breaking hydrodynamics with ${c}=0$, we find
\begin{align*}
\omega^2 \bm{\kappa}_\delta =q^2 \eta;,\quad
\omega^2 \bm{\kappa}_\epsilon =q^2 \eta_H,\quad 
\omega^2 \bm{\kappa}_{q}= \zeta, \quad 
\omega^2 \bm{\kappa}_{q\epsilon} = 0 .
\end{align*}
Thus thermal conductivities are directly related to viscosities. This has been also checked to be consistent with results based on the magnetically charged black hole solutions \cite{Hoyos:2015yna}.

\subsection{Hall transport for Skyrmions in insulating materials} \label{sec:WIInsulating}

By isolating the momentum independent terms in \eqref{MomWI2} proportional to $\delta^{jl}$ and $\epsilon^{jl}$, we arrive at the simple relations 
\begin{equation}\label{HallWI}
\begin{split}
& \omega^2 \bb{\kappa}^{(0)}_\delta = 0\;, \qquad 
\omega^2 \bb{\kappa}^{(0)}_\epsilon = c\;,
\end{split}
\end{equation} 
where the superscript $^{(0)}$ denotes the momentum independent part. Intuitively, the reason $ \bb{\kappa}^{(0)}_\delta $ vanishes and $ \bb{\kappa}^{(0)}_\epsilon $ does not is that Skyrmions are associated with spontaneously broken translation symmetry along with broken parity, whose imprints can only enter through the parity odd part of the conductivity at zero momentum. More precisely, the second identity predicts that the formation of a single Skyrmion results in the creation of a unit of thermal Hall conductivity $\bb{\kappa}^{(0)}_\epsilon$ in units of the quantized topological charge density. 
The frequency dependence is a consequence of the pole structure of the Goldstone boson that manifests itself in the retarded momentum correlator. In the presence of disorder, the behavior $ \bb{\kappa}^{(0)}_\epsilon = c /\omega^2 $ could in principle be lifted. However, recent numerical simulations have confirmed that Skyrmion motions are unaffected by impurities, in contrast to the case of domain walls \cite{Impurity}. The thermal Hall conductivity $\bb{\kappa}_\epsilon$ is dissipationless and exists even at zero temperature. While our Ward identity relations are valid at finite temperatures as well, measurements will be cleaner at very low temperatures, where additional dissipative contributions are suppressed. Another interpretation of eq. \eqref{HallWI} is that the Skyrmions carrying the thermal current propagate in an effective magnetic field given by the Skyrmion charge density $c_{ij}$, leading to a thermal Hall effect.  

For the momentum dependent terms in \eqref{MomWI2}, we obtain 
\begin{equation}\label{NeutralWIMomen}
\begin{split}
\omega^2 \bar{\bb{\kappa}}_\delta  
= q^2 \eta  \;, \quad
\omega^2 \bar{\bb{\kappa}}_\epsilon 
= q^2 \eta_H \;, \quad 
\omega^2 \bb{\kappa}_{q}  
= \zeta \;,
\end{split}
\end{equation} 
where the bar $~\bar{} ~$ indicates the nonconstant momentum dependent part; for example, 
$\bar{\bb{\kappa}}_\epsilon= \bb{\kappa}_\epsilon- \bb{\kappa}^{(0)}_\epsilon= q^2 \bb{\kappa}_\epsilon^{(2)} + q^4 \bb{\kappa}_\epsilon^{(4)} + \cdots $. Thus, thermal conductivities are directly connected to the viscosities of the system, as previously confirmed \cite{Hoyos:2015yna}. 
Furthermore, it follows from \eqref{MomWI2} that $ \bb{\kappa}_{q\epsilon} =0$. 

We can now describe a simple way to measure the Hall viscosity. Combining the second equations in \eqref{HallWI} and \eqref{NeutralWIMomen}, we get 
\begin{equation} \label{MeasuringHV}
\begin{split}
\eta_H = c \frac{ \bar{\bb{\kappa}}_\epsilon}{\bb{\kappa}^{(0)}_\epsilon} 
\to c \frac{ \bb{\kappa}_\epsilon^{(2)} }{\bb{\kappa}^{(0)}_\epsilon}    \;,
\end{split}
\end{equation} 
where we take the limit $q^2 \to 0$. Once the thermal Hall conductivity $\bb{\kappa}_\epsilon$ is measured as a function of $q^2$, the Hall viscosity is nothing but the Skyrmion density multiplied by the ratio between the slope and $\bb{\kappa}_\epsilon$-intercept $\bb{\kappa}_\epsilon(q^2=0)$. Note that this is only applicable in the presence of nonzero Skyrmion density.

\subsection{Generalization with a magnetic field and current }\label{sec:WIChargeB}

We generalize by allowing both a background magnetic field and a current. Then the conservation equation and momentum generators are modified.  
\begin{align}
\partial_\mu T^{\mu i}=B\epsilon^i_{\ j} J^j \;, \qquad 
T^{0j}_B = T^{0j}-\frac{B}{2}\epsilon^j_{\ n}x^n J^0 \;. 
\end{align} 
The second equation can be thought as a minimal coupling substitution, while the first one is from the general expression $ \partial_\mu T^{\mu\nu} = F^{\nu\rho}J_\rho$. Thus the momentum-momentum correlator $G^{0j0l} \sim \langle [T^{0j}, T^{0l}] \rangle$ on the left hand side of \ref{TranslationWI} will also include $G^{0j,m} \sim \langle [T^{0j}, J^m] \rangle$, $G^{n,0l} \sim \langle [J^n, T^{0l}]\rangle$, and $G^{nm} \sim \langle [J^n, J^m]\rangle$, while the only change in the right hand side is the addition of $B \langle J^0\rangle $. 

After some algebra, one obtains 
\begin{align}
\begin{split}
& \omega^2 G^{0j0l} -i\omega B\epsilon^j_{\ n}G^{n,0l}+i\omega B\epsilon^l_{\ m}G^{0j,m}
+B^2\epsilon^j_{\ n}\epsilon^l_{\ m} G^{nm}\\
&= - i\omega\epsilon^{jl}\left[ {c} -B\langle J^0\rangle \right]\delta (\vec x - {\vec x}' ) 
+\partial_n \partial_m' G^{njml}\;.
\end{split}
\end{align}
With index structures similar to \eqref{IndexStructures}, we re-express the equation using the notation for the thermoelectric conductivities $\bb{\alpha}\sim \langle[T,J]\rangle$ and $\bb{\alpha}^*\sim \langle[J,T]\rangle$ and electric conductivity $\bb{\sigma}\sim \langle[J,J]\rangle $ similar to thermal conductivity $\bb{\kappa}\sim \langle[T,T]\rangle $.
\begin{align}\label{ChargedWI}
\begin{split}
& \delta^{jl}  \big[\omega^2 \bb{\kappa}_\delta 
+ i\omega B \left(\bb{\alpha}_\epsilon+ {\bb{\alpha}}^*_\epsilon+ q^2 [ \bb{\alpha}_{q\epsilon}\!-\! {\bb{\alpha}}^*_{q\epsilon}]  \right) + B^2 \left( \bb{\sigma}_\delta + q^2 \bb{\sigma}_{q} \right) \big]  \\
&+{\epsilon^{jl} \big[\omega^2 \bb{\kappa}_\epsilon
	- i\omega B \big(\bb{\alpha}_\delta + {\bb{\alpha}}^*_\delta + q^2 [\bb{\alpha}_{q} + {\bb{\alpha}}^*_{q}]/2  \big) + B^2 \bb{\sigma}_\epsilon \big] } \\
&+ q^j q^l  \big[ 
\omega^2 \bb{\kappa}_{q} - 2i\omega B (\bb{\alpha}_{q\epsilon} -  {\bb{\alpha}}^*_{q\epsilon}) - 
B^2 \bb{\sigma}_{q} \big]   \\
&+ (\epsilon^{jo}q^l + \epsilon^{lo}q^j) q_o \big[ 
\omega^2 \bb{\kappa}_{q\epsilon} + i\omega B (\bb{\alpha}_{q} \!-\! {\bb{\alpha}}^*_{q})/2 - B^2 \bb{\sigma}_{q\epsilon} \big]  \\
&= \epsilon^{jl}  \big[{c} - B\rho + q^2 \eta_H \big] 
+ \delta^{jl} q^2  \eta + q^j q^l  \zeta   \;. 
\end{split}
\end{align}
This is a general formula in the presence of charge density $\langle J^0 \rangle = \rho$ and a magnetic field. Now allowing broken translational invariance, the following modifications appear after similar computations. 
\begin{align}
\eta_H \to \eta_H + \ell/2 \;, \qquad \zeta \to \zeta-i(p-BM)/\omega \;,
\end{align}
where $M$ is magnetization that can be defined  as $\langle J^{i} \rangle =\epsilon^{ik}\partial_k M$.

The momentum independent part of the Ward identities \eqref{ChargedWI} gives 
\begin{equation}\label{BHallWI}
\begin{split}
& \omega^2 \bb{\kappa}^{(0)}_\delta
+i\omega B (\bb{\alpha}^{(0)}_\epsilon+ {\bb{\alpha}}^{*(0)}_\epsilon) + B^2 \bb{\sigma}^{(0)}_\delta  
= 0 \;, \\
& \omega^2 \bb{\kappa}^{(0)}_\epsilon
-i\omega B (\bb{\alpha}^{(0)}_\delta + {\bb{\alpha}}^{*(0)}_\delta) + B^2 \bb{\sigma}^{(0)}_\epsilon 
= c - B\rho \;,
\end{split}
\end{equation}
where the superscript $^{(0)}$ denotes the momentum independent part. At non-zero momentum, there are four independent relations connecting viscosities and  conductivities as in the neutral case. In particular, the Hall viscosity is 
\begin{align} \label{BWIHallViscosity}
q^2 \eta_H = \omega^2 \bar{\bb{\kappa}}_\epsilon + B^2 \bar{\bb{\sigma}}_\epsilon 
-i\omega B \Big[\bar{\bb{\alpha}}_\delta + \bar{{\bb{\alpha}}}^*_\delta
+\frac{\bb{\alpha}_{q} + {\bb{\alpha}}_{q}^*}{2} \Big].
\end{align}
where the bar $~\bar{} ~$ indicates the non-constant momentum dependent part defined above.

\subsection{Ward identities for conductors at zero momentum} \label{sec:HallConductivity}

Let us describe a set of experiments that measure the electric Hall conductivity in the presence of Skyrmions. In an experimental paper \cite{HC1}, Lee et. al. have measured the electric Hall conductivity of the MnSi with varying magnetic fields for a wide range of fixed temperature that cover the A-Phase. 
What they observe is a unique step-function like excess of Hall conductivity only in the region of the A-Phase. Similar experiments have been done, both for varying magnetic fields with fixed temperature and for varying temperature with fixed magnetic fields, to observe a similar step-function like excess of the Hall conductivity in the presence of Skyrmions \cite{HC4}.  

How does this happen? Skyrmions are electrically neutral because they are made entirely out of spins. Then why do they display electric responses? In \cite{FerroCoupling2,Review} interactions between Skyrmions and conduction electrons are modeled by the ferromagnetic spin coupling. In the strong coupling limit, the spin wave function of the conduction electrons is identified with that of the localized spin $\vec n(x^\mu)$ of the Skyrmions. This limit is described by a tight binding model with Hund's rule coupling. There is an equivalent way to say this. The Skyrmionic spin configurations create an emergent magnetic field with magnitude $b=c/2$, where $c$ is the topological charge density of Skyrmions. This correspondence has been well known in $CP_N$ models. 

We model the effects of the interaction between the thermal and charge responses by modifying the parameters of the Ward identities. The Skyrmion charge density produces an emergent magnetic field $b=c/2$ \cite{Review}, which can change the dynamics of conduction electrons, similarly to $B$. For simplicity, we assume that the emergent magnetic field is homogeneous and constant, which is the case for all practical measurements. Due to the tight binding, the motion of the conduction electrons will also influence the thermal response of the Skyrmions. At vanishing momentum, by taking these effects into account, we get 
\begin{equation}\label{bBHallWI}
\begin{split}
\omega^2 \bb{\kappa}^{(0)}_\epsilon
-i\omega B_b (\bb{\alpha}^{(0)}_\delta + {\bb{\alpha}}^{*(0)}_\delta) + B_b^2 \bb{\sigma}^{(0)}_\epsilon  
= c_b - B_b\rho \;.
\end{split}
\end{equation}
This identity is of the same form as \eqref{BHallWI}, with the modification $B \to B_b \equiv B + b $ contributing to the charge response, and $c \to c_b \equiv c + c_{el}$ incorporating an additional contribution to the thermal response from the conduction electrons $c_{el}$, without changing the topological charge density. The quantities $c$ and $b$ are constant and independent of $B$, while $c_{el}$ (also measurable) is expected to be proportional to $B$ and depends on the strength of the binding. $b, c, c_{el}$ are expected to be readily identifiable experimentally. In particular, $b$ can be identified from a step-function-like signature in the Hall conductivity $\bb{\sigma}_\epsilon$ \cite{HC1,HC2,HC3,HC4}, as one passes into and out of a phase in which Skyrmions develop a finite density $c$. Such behavior will also confirm the presence of a nonzero density $c$, which will likewise produce a similar step-function-like contribution in the thermal Hall conductivity $\bb{\kappa}_\epsilon$ with an additional $B$-dependent $c_{el}$, by sweeping the magnetic field $B$ or the temperature $T$ independently.  

In the absence of ferromagnetic binding between the Skyrmion and conduction electron spins, the electric Hall conductivity would only pick up contributions from the conduction electrons, and $B_b$ would reduce to $B$. On the other hand, the thermal Hall conductivity would include  both contributions, $c$ and $c_{el}$, with the latter being independent of $B$.

\subsection{Hall viscosity and Skyrmions in conducting materials} \label{sec:HallViscosityConductor}

In this section we would like to propose a simple way to measure the Hall viscosity in conductors. Let us divide equation \eqref{BWIHallViscosity} by the second equation of \eqref{BHallWI} and substitute $ B \to B_b, c \to c_b$ as discussed in the previous section. We take the approximation $\omega/B_b \to 0$ and the limit $q^2 \to 0$, we obtain 
\begin{equation} \label{MeasuringHV2}
\begin{split}
\eta_H = (c_b - B_b \rho)~ \frac{ \bb{\sigma}_\epsilon^{(2)} }{\bb{\sigma}^{(0)}_\epsilon} \;.
\end{split}
\end{equation} 
Once the electric Hall conductivity $\bb{\sigma}_\epsilon$ is measured as a function of $q^2$, the Hall viscosity is nothing but the modified Skyrmion density $ c_b - B_b \rho$ multiplied by the ratio between the slope and $\bb{\sigma}_\epsilon$-intercept $\bb{\sigma}_\epsilon(q^2=0)$. In the opposite limit $B_b/\omega \to 0 $, $\eta_H$ reduces to 
\begin{equation} \label{MeasuringHV3}
\begin{split}
\eta_H = 
(c_b - B_b \rho)~ \frac{ \bb{\kappa}_\epsilon^{(2)} }{\bb{\kappa}^{(0)}_\epsilon}    \;.
\end{split}
\end{equation} 
Note that this identification of $\eta_H$ can also be applied to systems without Skyrmions, such as quantum Hall systems.

\subsection{Ward identities with angular momentum \& Hall viscosity} \label{WIAngularMomentum}

Finally, we discuss the Ward identities without translation invariance in the context of insulator. Generalization to include conductor has been done in \cite{Kim:2015qsa}. If the system of interest is not translationally invariant, there will be additional contributions to the Ward identity \eqref{NeutralWIMomen}; however, the zero momentum identity \eqref{HallWI} will be unmodified. A particularly interesting contribution of this type arises in parity-breaking systems exhibiting spontaneously generated angular momentum $\ell$ \cite{Toma1991,Liu:2012zm}, where the momentum generator can develop an expectation value 
\begin{equation}\label{Angmom}
\langle T^{0i} \rangle =\frac{1}{2}\epsilon^{ik}\partial_k\ell \;.
\end{equation}
In the absence of translation invariance, the two time derivatives $\partial_0 \partial'_{0}$ acting on $G^{0i,0j}(x^\mu, x'^\mu)$ pick up the contact term  $\frac{i}{2}(\partial'_0 - \partial_0) \big[ \delta(x^0\!- x'^0) \langle [T^{0i}(x^\mu), T^{0j}(x'^\mu)] \rangle \big]$ that appear in the second line of \eqref{GeneralWI}. The commutator yields a tensor similar to $\eta_H$ coming from the first term in the right side of \eqref{GeneralWI} \cite{Hoyos:2015yna}. As a result, $\eta_H$ in \eqref{NeutralWIMomen} is modified to $\eta_H  + \frac{\ell}{2}$. In such cases, a  coordinate space description  might be more convenient. Similarly, the inclusion of pressure $p$, another universal contribution, would replace $\zeta$ in the last relation of \eqref{NeutralWIMomen} by the combination $ \zeta -\frac{i}{\omega} p $.

Recently, Skyrmions have been observed in the insulating material Cu$_2$OSeO$_3$ \cite{Insulating1}, and various experiments regarding the Hall thermal conductivity and angular momentum have been carried out \cite{Rotation1,ThermalHC2,ThermalHC1}. For insulators, our Ward identity provides a simple relation among parity violating transport coefficients,  
\begin{align}\label{WIInsulator}
&\omega^2 \bb{\kappa}_\epsilon 
= c + \partial^2 \Big(\eta_H +\frac{\ell}{2} \Big)  \;,
\end{align} 
which is derived from eqs. \eqref{MomWI2} and \eqref{Angmom} in the absence of translation invariance. Recent experiments have successfully measured the Skyrmion density, thermal Hall conductivity and angular momentum in Skyrmion materials \cite{ThermalHC1}. Such measurements could in principle be used to infer the existence of Hall viscosity. \\

{\it Acknowledgements:} 
We are grateful to Alfred Shapere for helpful discussions and valuable comments on the draft. 
Various parts of this work have been presented over a period of time in numerous places, including the Berkeley Center for Theoretical Physics, Berkeley; Korea Institute for Advanced Study, Seoul; Great Lakes Strings Conference, Ann Arbor; SPOCK regional string meeting, Cincinnati. We thank to the members of the institutes and organizers of the meetings. We also thank to the organizers of the 62nd Annual Conference on Magnetism and Magnetic Materials (2017 MMM conference, Pittsburgh) for the invitation and an opportunity to put this into a written form.

\end{document}